\documentstyle[preprint,aps,prb,epsfig]{revtex}
\begin{document}

\title{Quantum decoherence in strongly correlated electron systems}

\author{Byung Gyu Chae}

\address{Basic Research Laboratory, ETRI, Daejeon 305-350, South Korea}

\maketitle{}

\begin{abstract}

  Complexity in strongly correlated electron systems
is analyzed by considering decoherence process between the localized state,
$|L$$>$ and the itinerant state, $|I$$>$.
The coherent superposition state of $\alpha |I$$> + \beta |L$$>$ decoheres to the pointer states
in the proximity of both extremes of the correlation
where the symmetry-breaking ground states of the charge pairing emerge.
For maximizing the entropy of the system,
the superconducting pairing and the spin density wave coexist within the uncertainty principle,
which invokes the metastable states as like pseudogap phase and electronic inhomogenity.

\end{abstract}

  Electrodynamics of strongly correlated electron systems has been known to be very complicate,\cite{1,2}
and especially, in cuprates exotic behaviors such as pseudogap state and superconductivity appear
with doping concentration.
Understanding of these properties is still evolving,\cite{3,4,5}
but satisfactory explanation has not existed yet
because it has a difficulty in applying the conventional perturbation analysis due to the strong interaction.
The dual behaviors of localization and itinerancy of charge particles in the correlated system
are general feature.
Therefore,
it is required for finding the consistent method to describe the duality
of charge dynamics.

  Previously, we reported on the basic concept of consistent description of dual effects
on charge dynamics simultaneously.\cite{6}
Based on the formalism of one-particle Green function,\cite{7}
each component of Hamiltonian for localization and itinerancy is extracted
from simply using the renormalization constant, $Z$.
The physical phenomena generated by two components are not independent each other and connected
through this factor.

  In this study, we note its role of quantum decoherence in the complexity of the correlated system.
As the correlated system approaches the side of the insulator or the Fermi liquid state,
decoherence process between localized and itinerant states undergoes,
where the system parameters as like low-dimensionality or the extremes of the correlation
play a role of its environment.
For maximizing the entropy of the system,
the symmetry-breaking ground states of both sides coexist to make the complexity.

  Spectral representation of charge dynamics of strongly correlated system clearly shows
two parts of spectral density consisting of quasiparticle and localized peaks.\cite{8}
The former peak represents the renormalized spectral weight of the quasiparticle
and the latter indicates the Hubbard bands extracted from the Hamiltonian
including on-site Coulomb repulsion in lattices.\cite{9,10}
Figure 1 shows the schematic diagram of spectral density divided into the itinerant state, $|I$$>$
and the localized state, $|L$$>$.
Splitting of the spectral density is typical phenomenon
appearing at the strong correlation.
Even though two states can be the integral quasi-states induced by the interaction among charge particles,
they should become the preferred states well describing the itinerant and the localized behaviors of electrodynamics.
Therefore, the wave function of charge particle is simply written as the coherent superposition of both states
as like two-level system.

\begin{equation} \label{eq:eps} |\Psi$$> = \alpha |I$$> + \beta |L$$> \end{equation}

where $\alpha$ and $\beta$ can be expressed by using the renormalization constant, $Z$.
Previously,
we extracted that the quasiparticle part of the Green function is weighted by the renormalization factor $Z$,
and the localized part by the factor $1 - Z$.\cite{6}
From this,
the complex numbers $\alpha$ and $\beta$ are represented as $Z^{1/2}$ and $(1-Z)^{1/2}$
except for the phase factor, respectively
and satisfied with normalization condition, $|\alpha|^2 + |\beta|^2 = 1$.

  Above two-level system has been known to evolve into the possible outcomes
at both extremes of the correlation strength,
that is, the Mott insulator or the Fermi liquid state.\cite{2}
We find that prior to arrival at the ultimate states,
the coherent superposition state can decohere to the pointer states.
Decoherence indicates the destruction of the coherence of the possible states,
where the density operator is a useful tool for describing this effect.\cite{11}
The most general density operator, $\rho$ is of the form representing a statistical mixture of pure states.

\begin{equation} \label{eq:eps} \rho = \sum_{i} g_{i} |\Psi_{i}$$><$$\Psi_{i}| \end{equation}

where the coefficient $g_{i}$ is the proportion of the ensemble being in the state $|\psi_{i}$$>$.
Considering the pure state $|\Psi$$>$ in eq. (1), the reduced density matrix can be simply expressed as below,

\begin{equation} \label{eq:eps} \rho = \alpha \alpha^* |I$$><$$I| + \alpha \beta^* |I$$><$$L|
                                       + \beta \alpha^* |L$$><$$I| + \beta \beta^* |L$$><$$L|
                                      \end{equation}

  The off-diagonal elements $\alpha \beta^*$ and $ \beta \alpha^*$ have the relative phase factor, $e^{i\theta}$.
In the many-particle system, the density operator is obtained from the ensemble average,
$\overline{\rho_{ij}}$ where the subscript are $\alpha$ and $\beta$.
The average cancels out these terms and thus,
the density matrix becomes diagonalized to vanish the interference effects.
In this system, decoherence process evolves at fast timescale.\cite{12}
The system will be in the state $|I$$><$$I|$ with probability $|\alpha|^2$ and in the state $|L$$><$$L|$
with probability $|\beta|^2$.

\begin{equation} \label{eq:eps} \rho = |\alpha|^2 |I$$><$$I| + |\beta|^2 |L$$><$$L|
                                      \end{equation}

  Decoherence arises from the interaction with environment,
which is a natural process of the loss of information from a system into the environment.
The system tends to progress in the direction of increasing entropy.
In strongly correlated electron system,
the system parameters such as low-dimensionality and the extremes of the correlation strength
play a role of its environment.
The variation of doping concentration or the lattice constant controls the correlation strength,
where the coherent superposition state can be forced into one of the diagonal eigenstates,
the collapse of the wave function.
Although decoherence cannot explain which of the possible outcomes emerges,
we find that between both extremes of the correlation, the physical behaviors
revealed by the localized and itinerant states appear.

  In interacting system,
the conventional Hamiltonian has the kinetic energy $\varepsilon_{k}$ and the interacting energy $V_{q}$.

\begin{equation} \label{eq:eps} H = \sum_{k} \varepsilon_{k} c^{\dagger}_{k} c_{k} +
                 \sum_{k,k',q} V_{q} c^{\dagger}_{k+q} c^{\dagger}_{k'-q} c_{k'} c_{k}  \end{equation}

  where $c^{\dagger}_{k}$ and $c_{k}$ are creation and annihilation operators, respectively.
Within the framework of Nambu-Gorkov formalism,\cite{13}
the symmetry-breaking ground states of the charge pairings of electron-hole and electron-electron are mixed.
The form of the interaction Hamiltonian for interacting system is as below.

\begin{equation} \label{eq:eps} H_{I} = \Delta_{SW} \sum_{k,\gamma, \gamma'} b^{\dagger}_{k+Q,\gamma} b_{k,\gamma'} +
                           \Delta_{SC} \sum_{k} a^{\dagger}_{k \uparrow} a^{\dagger}_{-k \downarrow} \end{equation}

  The gap parameter for the spin density wave $\Delta_{SW}$ and the charge pairing $\Delta_{SC}$ are expressed as
$U \sum_{k} \langle  b^{\dagger}_{k+Q,\gamma} b_{k,\gamma'} \rangle$
and $V \sum_{k} \langle a_{k \uparrow} a_{-k \downarrow} \rangle$, respectively.
We consider that the superconducting pairing comes from the itinerant state, $|I$$>$,
while the spin density wave with gap parameter $\Delta_{SW}$ comes from the localized state, $|L$$>$.\cite{14}

  The real systems, especially cuprates show the existence of the intermediate phase
consisting of the collective modes and the superconducting pairing
prior to arrival at the fully localized state or the Fermi liquid state.
This is the result from the decoherence process.
Both physical behaviors corresponding to the pointer states are revealed after decoherence,
and vice verse, they become an origin in decoherence.
Here, the spin density wave may have finite-range order 
because of two-dimensional behavior of charge carriers.
Microscopic origin of the superconducting pairing is not clear whether or not the bosonic glue mediates,\cite{15}
but the exchange force in many-body system may be probable one.

  We approach this problem from a macroscopic point of view, rather than a microscopic one.
It is instructive to investigate the effect of the change in the entropy of a system on its microscopic property.
As mentioned previously,
decoherence process progresses to increasing the entropy of a system.
This approach depends on the method of finding the probability distribution $p_{i}$
exhibiting the maximum value of the entropy $S$ when the constraints exist.
The entropy of a system is given by the equation.

\begin{equation} \label{eq:eps} S = - k_{\rm{B}} \sum_{i} p_{i} \log p_{i} \end{equation}

where $k_{\rm{B}}$ is the Boltzmann constant.
Using the method of Lagrange multipliers for the constraint, $\sum_{i} p_{i} E_{i} = \lambda$,
we obtain the general solution from the partial derivative of the quantity, $S - \beta \sum_{i} p_{i} E_{i}$.

\begin{equation} \label{eq:eps} p_i = \frac{\exp(-\beta E_{i})}{Z(\beta)} \end{equation}

  The partition function $Z$ is defined by $ Z(\beta) = \sum_{i} \exp(-\beta E_{i}) $.
Considering our model with two physical quantities of the spin wave and the superconducting pairing,
from a macroscopic viewpoint the entropy is of the simple form,
$ -k_{\rm{B}} p \log p - k_{\rm{B}} (1-p) \log (1-p) $.
In this case, the maximum entropy is achieved when the two states have equal probability,
that is, the same energy value.
We find that the existing value of the excitation energy of the spin wave can change that of the charge pairing,
and furthermore, induce the superconducting pairing itself for maximizing the entropy of the system.
Although the detailed analysis is required,
the probability of the superconducting pairing relates to the quantity,
$ \langle  b^{\dagger}_{k+Q,\gamma} b_{k,\gamma'} \rangle$ from in Eq. (8)
where the gap energy of the $\Delta_{SW}$ and $\Delta_{SC}$ is considered.

  Above description does not supply the solution of the microscopic origin in the charge pairing,
but there is a novelty that macroscopic parameters as like the entropy and information affects the microscopic quantity.
Namely,
a kind of the entropic energy becomes the parameter of forming the charge pairing.

  We find from the eq. (6) that both parts of the Hamiltonian does not rely on together
because they have independent eigenstates in time.
Therefore, both physical phenomena can be analyzed separately.
From the analysis of the expectation value,
the quasiparticle pairing and the collective mode are linked by the renormalization factor $Z$.
The superconducting gap equation is expressed as below,\cite{16}

\begin{equation} \label{eq:eps} \Delta_{SC} = 2\hbar \omega_{C} \exp(-\frac{1}{NV})  \end{equation}

  where $N$ is the density of states at the Fermi surface
and $\hbar\omega_{C}$ is the cutoff energy of the interacting parameter.
The cutoff energy relates to the entropic energy.
From this,
we simply extract that the critical temperature $T_{c}$ of the superconductivity
is proportional to the factors, $Z$ and $(1-Z)$.
Assuming that the factor $Z$ increases with the doping carrier,
the $T_{c}$ variation according to doping amount is similar to the experimental results.\cite{2}
We find that the variation of the kinetic energy rather than the potential energy strongly affects on
the superconducting state because of the superconducting pairing dependent
on the renormalization constant.
This is in agreement with the analysis of the spectral weight transfer of high-temperature superconductor
by using the optical spectroscopy.\cite{17}

  The respective phenomena generated from two states are affected by the uncertainty principle.
It is reasonable that the uncertainty principle in the two-level system is extracted
by using the Green function formalism.
The single-particle Green function is represented as
$G(xt,x't') = -i \langle \Psi_{\rm{o}}|T[\hat{\psi}(xt)\hat{\psi}^{\dagger}(x't')]|\Psi_{\rm{o}} \rangle$,
where the $\hat{\psi}$ and $\hat{\psi}^{\dagger}$ are the annihilation (creation) field operators.
\cite{7}
The expectation value of an operator $\widehat{Q}$ is formally represented as below,

\begin{equation} \label{eq:eps} \langle \widehat{Q} \rangle = -i \lim_{t \to t'^{+}} \lim_{x \to x'}
                                \langle \widehat{Q} G(xt,x't')\rangle  \end{equation}

  By using the wave function in Eq. (1),
the expectation value has the sum of two components of
$\langle \widehat{Q} \rangle_{I}$ in the itinerant state, $|I$$>$
and $\langle \widehat{Q} \rangle_{L}$ in the localized state, $|L$$>$.

\begin{equation} \label{eq:eps} \langle \widehat{Q} \rangle = \langle \widehat{Q} \rangle_{I}
                                + \langle \widehat{Q} \rangle_{L}
                                \end{equation}

  From this, the product, $\langle \Delta Q \rangle \langle \Delta P \rangle$
of the uncertainties of conjugate variables, $\Delta Q$ and $\Delta P$ has three parts.
Two parts are related to the quantities for the itinerant and localized state, respectively.
The rest is the product of the respective variable in each state
and in this case, the uncertainty principle for position and momentum is given by,

\begin{equation} \label{eq:eps} \langle \Delta x \rangle_{I} \langle \Delta p \rangle_{L}
                                \geq Z(1-Z)\frac{\hbar}{2}   \end{equation}

  where the factors $Z$ or $1-Z$ are calculated from the expectation value by using
the Green function formalism because as previously described,
the Green functions of itinerant and localized states are expressed as having renormalization
factors, $Z$ or $1-Z$.
At the ultimate region, that is,
in case $Z$ is zero or one, this type of the uncertainty relation disappears.
Above uncertainty relation indicates that the physical phenomena generated by two possible states
are not independent together.
This invokes the system to have electronic inhomogenity,
where the collective modes and superconducting pairing are generated
within the extent of the uncertainty relation and coexist in space.
Therefore, the electronic inhomogenity is an intrinsic property appeared by the basic quantum effects
in the many-body system with the strong correlation.

  We find that ultimately, the tendency of the system to achieving the maximum entropy
rather than other features such as the inhomogeneous charge distribution
with the minimum energy causes the complexity of the system.
Here, we can estimate the extent of the spatial distribution, $\delta x$ of each physical phenomenon.

  \begin{equation} \label{eq:eps} \delta x \simeq \frac{\hbar p_{F}}{m \Delta}   \end{equation}

  where the $p_{F}$ and $m$ are the Fermi wave vector and mass of the carrier.
Considering the observed energy scale of about several decades meV of the density wave,
the spatial scale of them extends over several decades \AA.

  In conclusion,
the complexity of strongly correlated electron system can be understood
by decoherence process.
We find that decoherence process plays a key role in analyzing pseudogap phase and superconductivity.

\begin{figure}
\vspace{2.0cm}
\centerline{\epsfysize=7cm\epsfxsize=9cm\epsfbox{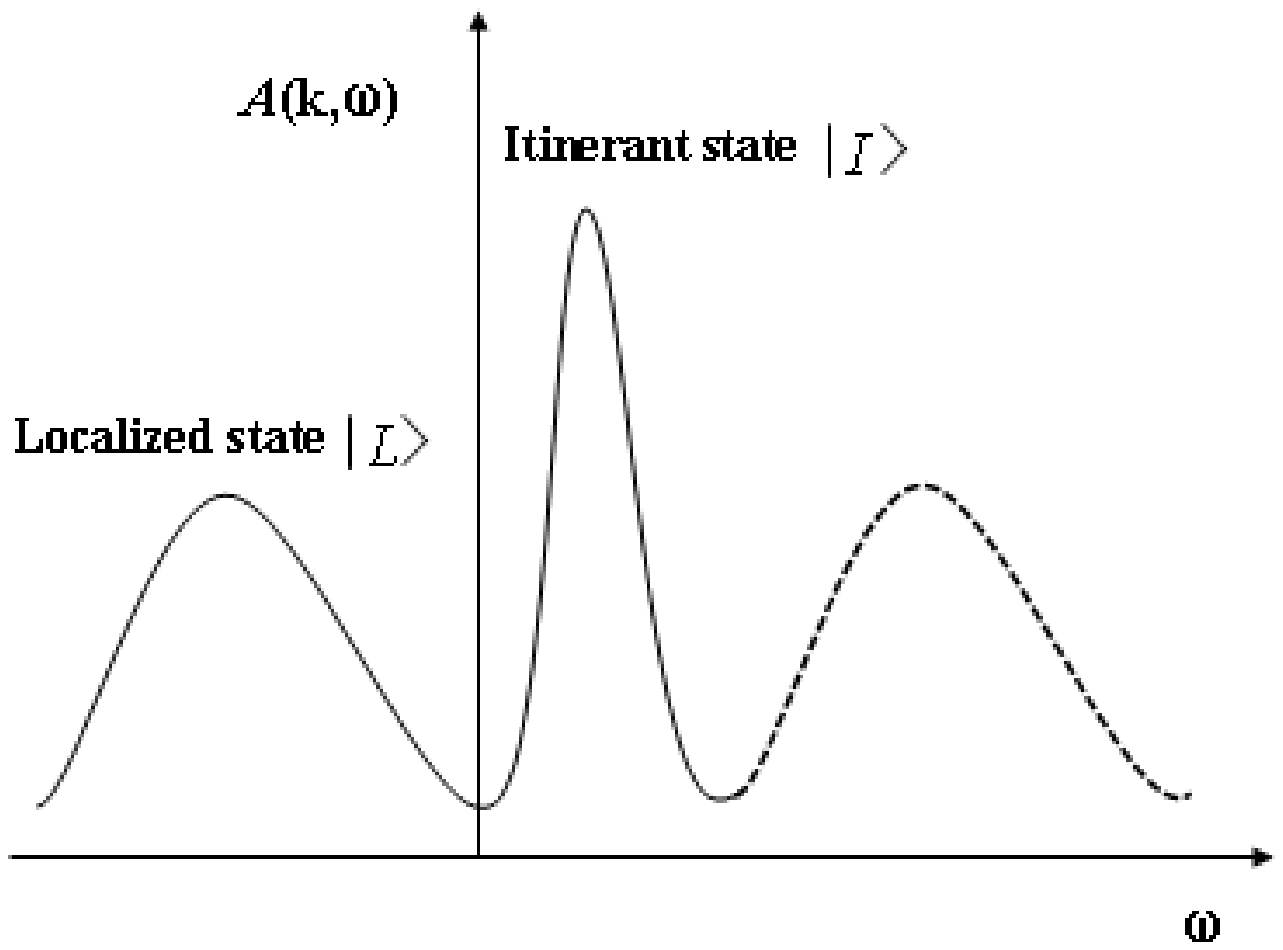}}
\vspace{0.0cm}
\caption{The schematic diagram of spectral density separated into the itinerant state, $|I$$>$
and the localized state, $|L$$>$}
\label{f1}
\end{figure}

\end{document}